\documentclass[conference]{ieeetran}
\IEEEoverridecommandlockouts
\usepackage{cite}
\usepackage{amsmath,amssymb,amsfonts}
\usepackage{algorithmic}
\usepackage{graphicx}
\usepackage{textcomp}
\usepackage{xcolor}
\def\BibTeX{{\rm B\kern-.05em{\sc i\kern-.025em b}\kern-.08em
    T\kern-.1667em\lower.7ex\hbox{E}\kern-.125emX}}

\usepackage{hyperref}
\hypersetup{
    colorlinks=true,
    linkcolor=blue,
    filecolor=blue,      
    urlcolor=blue,
    pdftitle={Overleaf Example},
    pdfpagemode=FullScreen,
    }
\begin{document}

\title{The Impact of COVID-19 on Chronic Pain: Multidimensional Clustering Reveals Deep Insights into Spinal Cord Stimulation Patients \\
\footnotesize
\thanks{\textsuperscript{*} Equal~contrib.~~\textsuperscript{+} Corresp.~author~~ \textsuperscript{++} Current~Affil.~Xilis,~Inc }
}
\author{\IEEEauthorblockN{Sara E. Berger \textsuperscript{*}}
\IEEEauthorblockA{\textit{IBM Research} \\
sara.e.berger@ibm.com}
\and
\IEEEauthorblockN{Carla Agurto \textsuperscript{*}}
\IEEEauthorblockA{\textit{IBM Research} \\
carla.agurto@ibm.com}
\and
\IEEEauthorblockN{Guillermo A. Cecchi \textsuperscript{+}}
\IEEEauthorblockA{\textit{IBM Research} \\
gcecchi@us.ibm.com}
\and
\IEEEauthorblockN{Elif Eyigoz}
\IEEEauthorblockA{\textit{IBM Research} \\
ekeyigoz@us.ibm.com}
\and
\IEEEauthorblockN{Brad Hershey}
\IEEEauthorblockA{\textit{Boston Scientific} \\
Brad.Hershey@bsci.com}
\and
\IEEEauthorblockN{Kristen Lechleiter \textsuperscript{++}}
\IEEEauthorblockA{\textit{Boston Scientific} \\
kristen.lechleiter@xilis.net}
\and
\IEEEauthorblockN{NAVITAS and ENVISION Studies Physician Author Group\textsuperscript{**}}
\IEEEauthorblockA{\textit{Boston Scientific} \\
}
\and
\IEEEauthorblockN{Dat Huynh}
\IEEEauthorblockA{\textit{Boston Scientific} \\
Dat.Huynh@bsci.com}
\and
\IEEEauthorblockN{Matt McDonald}
\IEEEauthorblockA{\textit{Boston Scientific} \\
matthew.mcdonald@bsci.com}
\and
\IEEEauthorblockN{Jeffrey L. Rogers}
\IEEEauthorblockA{\textit{IBM Research} \\
jeffrogers@us.ibm.com}
}

\maketitle

\begin{abstract}
The emergence of COVID-19 offered a unique opportunity to study chronic pain patients as they responded to sudden changes in social environments, increased community stress, and reduced access to care. We report findings from n=70 Spinal Cord Stimulation (SCS) patients before and during initial pandemic stages resulting from advances in home monitoring and artificial intelligence that produced novel insights despite pandemic-related disruptions. From a multi-dimensional array of frequently monitored signals—including mobility, sleep, voice, and psychological assessments—we found that while the overall patient cohort appeared unaffected by the pandemic onset, patients had significantly different individual experiences. Three distinct patient responses (sub-cohorts) were revealed, those with: worsened pain, reduced activities, or improved quality-of-life. Remarkably, none of the specific measures by themselves were significantly affected; instead, it was their synergy that exposed the effects elicited by the pandemic onset. Partial correlations illustrating linked dimensions by sub-cohort during the pandemic and those associations were different for each sub-cohort before COVID-19, suggesting that daily at-home tele-monitoring of chronic conditions may reveal novel patient types. This work highlights the opportunities afforded by applying modern analytic techniques to more holistic and longitudinal patient outcomes, which might aid clinicians in making more informed treatment decisions in the future.
\end{abstract}


\section{Introduction}
The limitations of traditional patient care have been starkly exposed by the COVID-19 pandemic. There have been significant disruptions to patient care, treatment paradigms, and on-going clinical studies \cite{b1}. Patients with chronic pain have been particularly vulnerable, as the pandemic interrupted many of their daily routines, pain management plans, and analgesic therapies (including postponed medical office visits, canceled therapy sessions, delayed or limited analgesic treatments, and cessation of elective intervention procedures) \cite{b2}, as well as halted social or extracurricular activities which often serve as vital coping mechanisms for individuals \cite{b3}. Many pain patients were also more at risk of becoming seriously ill due to underlying concurrent health issues and concomitant medications which may affect respiratory and/or immune responses \cite{b3}. All of these disturbances may contribute to increased pain intensity, worsening disability, and deteriorating mood in a population which already has significant mental and physical health co-morbidities. 

Because chronic pain is such a complex, multidimensional experience that significantly impacts individuals’ mental and physical well-being, there is a need to capture more aspects of patients’ unique experience with additional metrics beyond the previous gold standard of pain intensity. Moreover, the sporadic access to clinicians and the reluctance of patients to visit healthcare centers during the pandemic shows how in-person interventions were constrained to major events, pushing to the forefront the need and opportunity for more remote interventions. Ecological momentary assessments (EMAs) are one such tool that allows measurement of various aspects of an experience repeatedly in real-time in different environments.  Coupled with a digital health ecosystem that not only collects self-reported EMAs, but also sensor-based EMAs and implanted medical device data, it is possible to obtain a rich, multi-dimensional, highly frequent, and minimally intrusive representation of patients as they go about their daily lives. They can help us not only evaluate patients when they cannot come to laboratory or clinical visits, but also obtain a more holistic and multi-faceted picture of a patient’s experience outside of the physical boundaries of a doctor’s office or research lab. To this point, we have already witnessed a large effort to make such methods more mainstream during COVID-19 for a gamut of clinical and non-clinical use cases \cite{b4,b5,b6,b7}. 

On-going clinical research of spinal cord stimulation (SCS) patients using a digital health ecosystem presented a unique opportunity to evaluate the extent to which the pandemic impacted on-going chronic pain experiences and quality-of-life (QoL). In particular, we hypothesized that under the additional economic, social, and healthcare-access stressors provoked by the COVID pandemic: (1) a significant diversity in pain patients’ responses would be exposed, (2) these responses would not be restricted to or explained by canonical assessments based on reported pain intensity scores, and (3) in-home continuous evaluation of multiple self-reported and sensor-based assessment streams would reveal strong associations (and concomitant dissociations) that may eventually inform the existence of a multi-modal chronic pain state indicator. To test these hypotheses, we analyzed self-reported ratings, physiological measurements from a smart watch and sleep sensor, and speech recordings collected before and during COVID from patients engaged in these studies for the treatment of chronic low back and/or leg pain with SCS.

\section{Methods}

\subsection{Study Data Collection}
We collected data through multiple ongoing longitudinal, observational, and prospective clinical studies (NAVITAS and ENVISION Studies, Clinicaltrials.gov ID:  NCT03240588) involving up to 1700 chronic low back and/or leg pain patients who are or were candidates for SCS treatment (Boston Scientific, Valencia, CA). Subjects were enrolled at multiple sites in the United States if they were planning to receive or had already received an SCS trial or implant system. Additionally, subjects may have been previously enrolled in the RELIEF study (Clinicaltrials.gov ID:  NCT01719055).  This analysis focuses on those subjects enrolled in the ENVISION study who met the analysis criteria, however the dataset may have included patients from any of the studies above if the subject was enrolled in more than one study concomitantly.

The ENVISION study collects a variety of self-reported, psychological, physiological, and other measures both in-clinic via cross-sectional visits and longitudinally at-home, via a custom digital health ecosystem, for up to 12 – 24 months. Outcome measures include pain, standardized questionnaire scores, mood/emotion, voice, sleep quality/quantity, cardiac function, wrist worn actigraphy, medication use, spinal fluoroscopic imaging, SCS parameters/usage, functional fitness, treatment satisfaction, and quality-of-life reports. At-home data were collected using a custom-built digital health ecosystem (Boston Scientific, Valencia, CA) connected to wearable accelerometers (Galaxy Watch S2, Samsung USA, Menlo Park, CA with custom watch application, Boston Scientific, Valencia, CA), ambient sleep sensors (Emfit QS, EMFIT, San Marcos, TX), and the SCS system (Spectra WaveWriterTM/Precision SpectraTM, Boston Scientific, Valencia, CA).  

\subsection{Data Periods}
This analysis compares a subset of collected EMA \cite{b8} data from eligible ENVISION subjects early in the COVID-19 pandemic (when the uncertainty and changes were expected to be the most significant) to the same assessments from the same patients prior to the pandemic. Since this clinical cohort is exclusively in the USA, we used the dates of the COVID-19 national emergency to define pandemic period and pre-pandemic (baseline) period that are used for the majority of comparisons. Specifically, the US government publicly declared a national emergency 13 March 2020 \footnote{https://www.whitehouse.gov/briefing-room/presidential-actions/2022/02/18/notice-on-the-continuation-of-the-national-emergency-concerning-the-coronavirus-disease-2019-covid-19-pandemic-2/)}; however, the national emergency technically began on 1 March 2020 and news coverage was actively reporting on the pandemic before this dates. Therefore, we used data collected during the 6 weeks between 6 March 2020 (half-way point) and 17 April 2020 for the Pandemic Period. We then defined a 5-week Gap Period (1st February 2020 - 5 March 2020) to account for growing awareness in the patient cohort to the emerging pandemic. The 6 week-period prior to this gap was used as the Baseline Period (20 December 2019 – 31st January 2020) for comparing the patients to the Pandemic Period. The Baseline period was chosen to maximize patient sample size while still accounting for other time effects (e.g., being close in time to the Pandemic Period and having the same analysis window length of 6-weeks duration). Patients were given a set of targeted questions (COVID Questions) during the final week of the study-defined Pandemic Period (10 April 2020 - 17 April 2020).

\subsection{Datasets and Analyses}
To be eligible for analysis, patients had to be actively enrolled in the ENVISION study through the Baseline and Pandemic Periods, have responded to at-home study questions during the Baseline Period and the COVID Questions, and have the same treatment status during the Baseline and Pandemic periods (i.e., if they were pre-implant during the Baseline Period, they were pre-implant during the Pandemic Period). The EMAs evaluated here included a subset of in-clinic questionnaires, self-reported ratings, physiological measurements, and voice recordings.  

In-clinic questionnaires were collected at baseline, 1-month, 3-months, and 12 months post enrollment, with an optional extension to 24 months. Here, we analyze Beck Depression Inventory (BDI-II), Fear Avoidance Beliefs Questionnaire (FABQ), Oswestry Disability Index Version 2.1a (ODI v2.1a), and Pain Catastrophizing Scale (PCS) from either the most recent visit attended within the Baseline Period or the most recent in-clinic visit prior to Baseline (enrollment, 1-month enrollment, or 3-months, whichever was closest to the start of the Baseline period).  Patients were asked to provide self-reported ratings at home on their phones using the digital ecosystem at least once or up to twice a day.  This analysis focuses on pain intensity (0.0 – 10.0 numeral rating slider), mood (1 – 5 stars), sleep quality (1 – 5 stars) and duration (hours), medication intake (“I didn’t use any”, “less than usual”, “same as usual”, “more than usual” for opioids; non-opioid prescription medications with indications for pain, over-the-counter medications, and prescribed sleep medications were reported separately), and activities of daily living (list of 13 activities that can be summarized as self-care, exercising, commuting, and socializing). Physiological features extracted from the watch, worn daily, included step counts, activity intensity, and effective mobility \cite{b9} (an engineered metric derived from the smartwatch sensors that quantifies both intensity and duration of activity). Overall sleep score, duration to sleep onset (the transition time from wakefulness into the first stages of sleep), toss and turn counts, number of awakenings, and lowest heart rate information achieved during sleep (precalculated by EMFIT, San Marcos, TX) \cite{b10,b11,b12} were also analyzed for all patients who chose to use the optional ambient sleep sensor. Finally, voice recordings were collected using the digital ecosystem to administer speech prompts approximately once a week (with an option to submit more often if they wanted). Extracted acoustic and content (psycholinguistic) features used were those that have been previously found in the literature to be associated either with pain \cite{b13}, such as pitch variation or the second formant (frequency response of the acoustic resonance of the human vocal tract), or with emotional state \cite{b14}, such as association with emotional concepts or overall sentiment \cite{b15}. To analyze the syntactic and semantic content of patients’ speech, voice recordings were first sent to a transcription service (TranscribeMe, Oakland, CA, www.transcribeme.com) for manual transcription. Transcribed files were then fed into a number of programs, including a slot grammar parser \cite{b16} to extract and normalize syntax-related content features and a cloud-based service that uses deep learning to extract metadata from text such as entities, keywords, categories, sentiment, emotion, relations, and syntax natural language understanding tools (Watson Natural Language Understanding, IBM, Armonk, NY, https://www.ibm.com/cloud/watson-natural-language-understanding) \cite{b17} to quantify sentiment and emotion-related content features. Analysis of variance (Kruskal Wallis and two-way ANOVA) and chi-squared tests were used to test for differences in patient demographics. The Wilcoxon signed rank test was used to compare EMAs between time periods for the analysis population. 

Subgroups were identified based on changes in their self-reported ratings between periods (Baseline-Pandemic) using a k-means clustering approach. The optimal number of clusters was determined based on the data resolution, sample size, interpretation potential, and stability as established with a Silhouette analysis, which estimates cluster separability function $S$ by comparing intra- and inter-cluster average distances. To identify the optimal number of clusters, we computed the drop-in separability as a function of increasing number of clusters, $\Delta S(n) = S(n)-S(n-1)$, and chose to select the cluster size such that adding one more cluster would result in the largest separability drop defined as $n^{\star} = \{n: \max(\Delta S(n+1)) \}$ over 1,000 repetitions).  Multidimensional Scaling (MDS) was used to preserve the similarity of the high-dimensional EMA data while projecting it into a new space with lower dimensions. Metric stress (normalized sum of squares) of the cosine distances was used to optimize the MDS. Partial correlations were used to test for statistically significant differences in the sub-cohorts’ EMAs between Baseline and Pandemic periods. For embedding the different periods’ partial correlations, we used Procrustes similarity between the matrices, which assumes that translation, rotation, and scaling preserves the identity of matrices. To estimate an index of robustness, we created replicas of the partial correlation for each period by repeatedly subsampling the cohorts independently. Mann Whitney and Wilcoxon signed rank tests were used to test for differences in physiological measurements obtained from the objective sensors. Differences between cohorts in in-clinic questionnaires and COVID Survey responses were determined using one-way ANOVAs and Kruskal-Wallis tests.  Extracted acoustic and content (psycholinguistic) features from voice data were evaluated using Kolmogorov-Smirnov tests, Wilcoxon signed rank tests, and interquartile range.

\section{Results}

\subsection{Study and Analysis Population }
\begin{figure}[htbp]
\centerline{\includegraphics[scale=0.7]{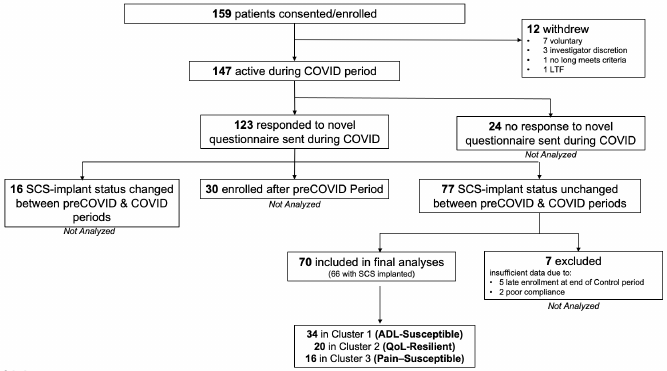}}
\caption{{\bf Patient population.} CONSORT diagram outlining entire cohort, from enrollment to analyses.}
\label{fig:population}
\end{figure}

We evaluated 159 patients enrolled in ENVISION for inclusion in this analysis (Figure \ref{fig:population}). Seventy of these subjects met the analysis criteria: 61.4\% female, 60 ± 9.4 years mean age, 15.1 ± 10.7 years mean duration of chronic pain, and 171.8 ± 58.7 days mean ENVISION study follow-up duration (Figure \ref{table1}).  All 70 subjects were candidates for SCS treatment of chronic pain in the low back (98.6\%), unilateral leg (41.4\%), and/or bilateral leg (42.9\%), and 94.3\% (n=66/70) received SCS treatment before the Pre-Baseline period.

\subsection{Self-reported ratings across the entire patient population were unchanged during the Pandemic Period}
\begin{figure}[htbp]
\centerline{\includegraphics[scale=0.7]{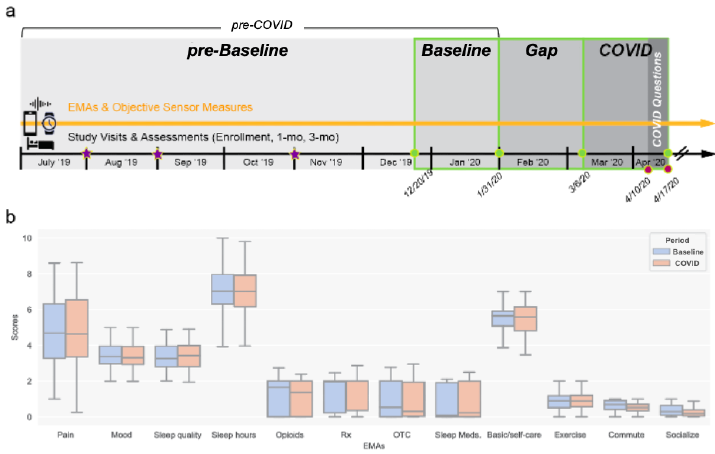}}
\caption{ {\bf Defining periods and comparing EMAs.} (a) Daily electronic self-reported ecological momentary assessments (EMAs), semi-continuous wrist-worn and sleep sensor signals, and weekly voice recordings (orange line), were compared between the Baseline period (20 December 2019 to 31 January 2020) and Pandemic period (6 March 2020 to 17 April 2020). Data were collected from each subject beginning at study enrollment, beginning with the initial patient in July 2019. A battery of standardized questionnaires was collected at enrollment and at any subsequent visits (stars and black line) prior to the COVID-19 pandemic; a set of 5 questions was also administered via the digital health ecosystem app for 1 week (COVID-Questions, red dots).  A Pre-Baseline Period captured in-clinic questionnaire data, enrollment information, and additional voice responses which had taken place prior to the pandemic. The Gap period between Baseline and COVID was defined as a buffer for COVID-19 spread and pandemic awareness. Periods indicated with green dots and boxes. (b) Box-and-whisker plots show averaged ratings of 12 EMAs across the entire Baseline period (blue) and Pandemic period (red) for n=70 patients; Wilcoxon tests revealed no differences between periods at a cohort-level, with the exception of commuting and socializing, which were lower during COVID}.
\label{fig:periods}
\end{figure}

With the exception of commuting and socializing, which were impacted at-large by COVID, no distinctions in self-reported ratings between periods were detected at a population-level (Figure \ref{fig:periods}b), Wilcoxon tests non-significant) in the at-home data. We reasoned that patients likely diverged in how they coped physically and mentally during COVID, and that these individual differences may be diluted or obstructed by focusing on the entire cohort or on specific EMAs as opposed to their interactions. This was confirmed by analyzing the covariance of ratings across time, where we repeated subsampled 50 embeddings of the ratings in each period and randomly selected 40 subjects in each iteration to compute a representation of each of the 4 periods, resulting in 50 different embeddings for each of the 4 periods.  The embeddings were stable for each period, and moreover, the first factor/dimension clearly separated the COVID period ratings from those in Baseline and Gap.  This demonstrated that the patients’ experiences during COVID were markedly and robustly distinct with respect to their previous reports and experiences.

\subsection{Clustering reveals linked factors in 3 sub-cohorts: Activities of Daily Living Susceptible, Pain Susceptible, and Quality of Life (QoL) Resilient}
\begin{figure}[htbp]
\centerline{\includegraphics[scale=0.8]{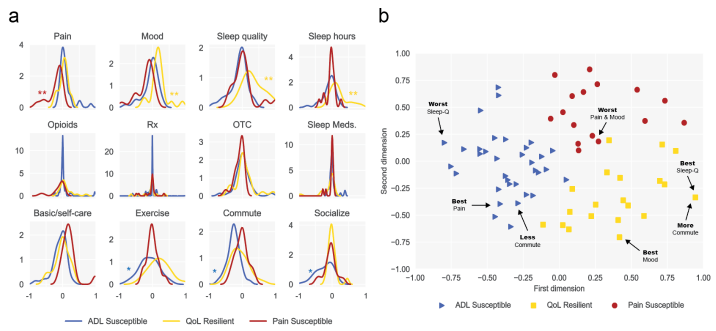}}
\caption{ {\bf Clustering results of the multivariate analysis.} K-means clustering produced 3 distinct sub-cohorts from 70 patients: 34 ADL Susceptible (blue); 20 QoL Resilient (yellow); 16 Pain Susceptible (red). Density plots (a) for changes in each of the self-reported variables between periods (Baseline, Pandemic). X-axis values above 0 indicate improvement during COVID period; below zero is worsening during COVID.  * indicates if the curve is different than at least one curve (KS test, p-value $<$ 0.05), ** indicates if the curve is different than the two other curves (KS test, p-value $<$ 0.05). (b) Reduced dimensionality representation of changes in self-reports, identified by cluster identity. Arrows indicate the subjects with the best or worst changes between Baseline period and Pandemic period across various aspects of the patients’ experiences – pain, QoL, and activity. Abbreviations: Sleep-Q (sleep quality); Sleep-Hrs (sleep hours); Rx (non-opioid prescription medications for pain indication); OTC (over-the-counter medications); Sleep-Ms (sleep medications); Basic/Self-care (basic activities and self-care activities, including lying down, sitting, standing, bathing, dressing, eating, \& cooking); Exercise (includes exercise, housework, and yard-work); Commute (includes all forms of traveling).}
\label{fig:clustering}
\end{figure}

\begin{figure}[htbp]
\centerline{\includegraphics[scale=0.6]{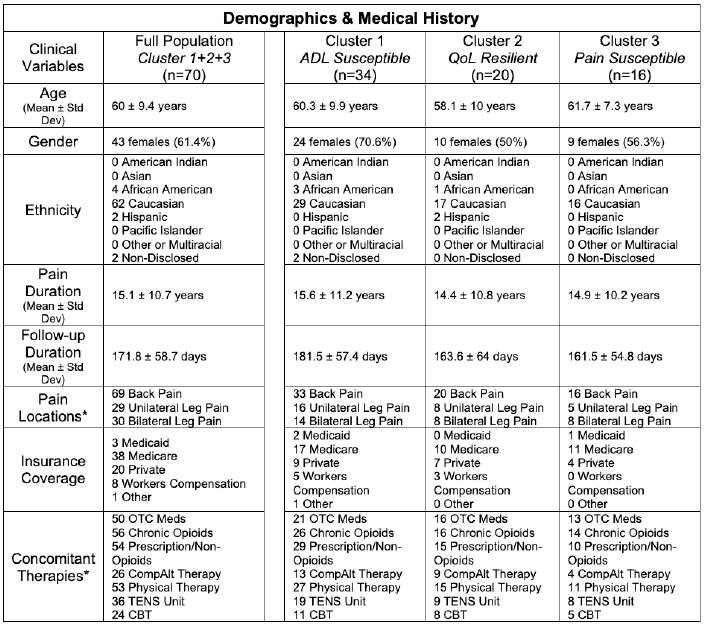}}
\caption{ {\bf Summary of key demographic and enrollment clinical variables.} Left side provides summary for analysis cohort and right side is same data, divided by K-means cluster assignment as in Figure \ref{fig:clustering}.  *indicates that these variables are non-exclusive (patients may fall into more than one category). There were no significant differences between cluster subcohorts in any of these measures (see supplementary material for statistics). Abbreviations: OTC – Over the Counter; CompAlt (complementary and alternative therapy); CBT – Cognitive Behavioral Therapy.}
\label{table1}
\end{figure}

Using changes in self-ratings between periods (Baseline-Pandemic), 3 clusters were determined (Figure \ref{fig:clustering}):  Cluster (sub-cohort) 1 was defined primarily by worsened activity levels during COVID, including reduced exercising, commuting, and socializing (ADL Susceptible, n = 34); Cluster 2 was defined by improved QoL measures including better mood, improved sleep, and reduced medication intake (QoL Resilient, n = 20); and Cluster 3 was defined primarily by worsening pain intensity during COVID (Pain Susceptible, n = 16). Associated differences in EMA changes between clusters can be found in Figure \ref{fig:clustering}a, showing that clustering separates groups with several simultaneous statistically significant differences.  Importantly, the groups did not differ in their demographics including potential confounding factors such as duration of pain or pain locations (Figure \ref{table1}). In Figure \ref{fig:clustering}b, we show the result of applying Multidimensional Scaling (MDS) to the rating changes, along with the cluster identity of the samples as computed by k-means. It can be seen how the projected samples are located such that both left-right and down-up directions correspond to worse-to-better pain and QoL. This two-dimensional mapping separates the three sub-cohorts very well, suggesting that the correlated EMA changes (12 dimensions) induced by the Baseline-to-Pandemic transition can be represented by a reduced number of independent dimensions.

\begin{figure}[htbp]
\centerline{\includegraphics[scale=0.8]{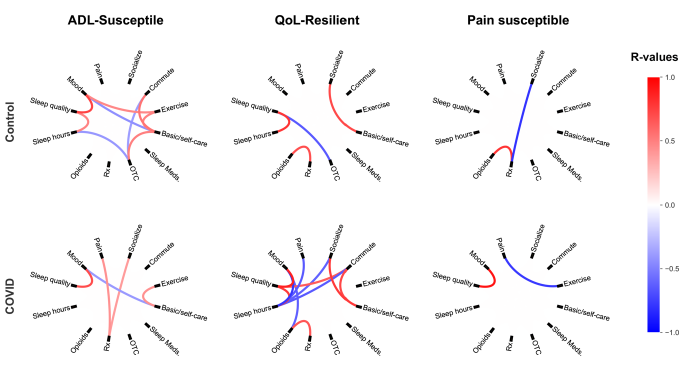}}
\caption{{\bf Partial correlations for self-reported variables.} Spider plots show conditional relationships (partial correlations) between self-report ratings across the 3 sub-cohorts (X-axis) for both pre-COVID and COVID periods (Y-axis); differences between plots can be visually appreciated. Nodes in plots indicate each of the 12 self-reported ratings (or combination of ratings). Sample sizes are identical to those in Figure 2.  Lines indicate partial correlations between self-report variables (correlations after accounting for all other relationships between variables); line color is the direction of correlation (positive = red; negative = blue) and line thickness/saturation indicates strength of correlation (larger r-values = thicker, darker lines; smaller r-values = thinner, lighter lines). Only the strongest (significant or trending) correlations are shown; if no line is present, variables were conditionally independent, or correlations were not statistically strong (thresholded at p = 0.075). Abbreviations: Sleep-Q (sleep quality); Sleep-Hrs (sleep hours); Rx (non-opioid prescription medications for pain indication); OTC (over-the-counter medications); Sleep-Ms (sleep medications); B-SelfCare (basic activities and self-care activities, including lying down, sitting, standing, bathing, dressing, eating, \& cooking); Exercise (includes exercise, housework, and yard-work); Commute (includes all forms of traveling).}
\label{fig:partialcorr}
\end{figure}

In contrast to the univariate analysis, we found statistically significant differences in the sub-cohorts when looking at partial correlations (pCorr) for 12 EMAs between Baseline and Pandemic periods:  pain, mood, sleep quality, sleep hours, opioid medications, non-opioid prescription medications indicated for pain, over-the-counter medications, sleep medications, basic self-care (including lying down, sitting, standing, bathing, dressing, eating and cooking), exercise (including exercising, housework, and yard-work); commute (includes all forms of traveling), and socializing. Figure \ref{fig:partialcorr} shows the linked factors (R-value color-coded, $p<0.075$) estimated for the 3 sub-cohorts in the two periods. The absence of a link between two EMAs in pCorr means that the corresponding ratings do not have any meaningful association after possible confounds are removed (i.e., they are conditionally independent) and as such do not directly influence each other. Note that in the Baseline period, pain intensity appears to be conditionally independent from all other ratings for the 3 sub-cohorts, an effect that changes during COVID, which shows different associations for pain in each group. In other words, our hypothesis that the stress introduced by COVID would impact perceived pain intensity is supported by the Pandemic period correlations with stronger dependencies on pain from the other linked factors. Another striking feature is that the 3 sub-cohorts show a strong association between sleep quality and mood during the Pandemic period, something only seen in the ADL-Susceptible sub-cohort during Baseline. 

\subsection{Subcohorts differ in their physiological data collected from sensors}
\begin{figure}[htbp]
\centerline{\includegraphics[scale=0.8]{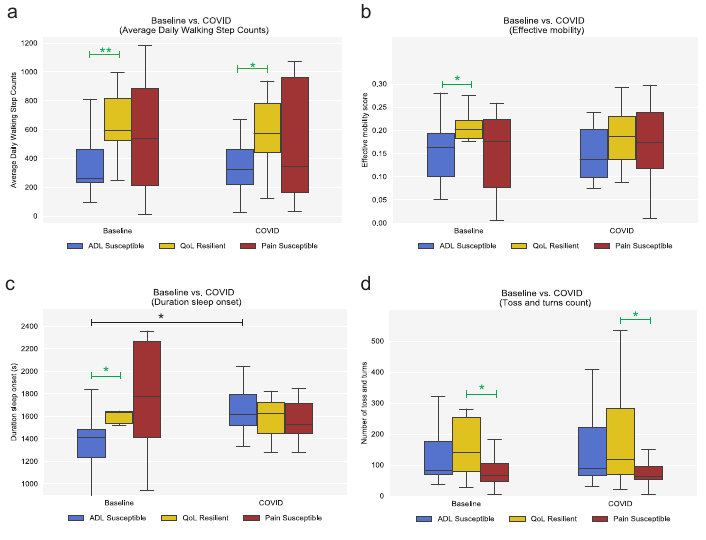}}
\caption{{\bf Boxplots of features from sensor-based measurements.} We collected objective (non-self-reported) features from patients, including step counts and effective mobility from a smart watch and a variety of measures from an under-the-mattress sleep sensor. Horizontal bars indicate significant differences using Mann-Whitney (green) and Wilcoxon paired test (black). * and ** at the top of the bars indicates p$<$0.05 and p$<$0.01, respectively. (a) Smart watch running step-counts compared between groups for both Baseline period and Pandemic period; step counts were totaled per day and averaged across each period (N = 16, 15 and 8 for ADL Susceptible, QoL Resilient and Pain Susceptible sub-cohorts, respectively). (b) Effective mobility scores derived from the watch accelerometer were averaged across each period and compared between sub-cohorts during the Baseline period or Pandemic period (N = 17, 16 and 8 for ADL Susceptible, QoL Resilient and Pain Susceptible sub-cohorts, respectively). (c) Duration sleep onset in seconds and (d) Toss and turns counts obtained from sleep sensors were averaged across each period and compared between each sub-cohort (N = 18, 10 and 8 for ADL Susceptible, QoL Resilient and Pain Susceptible sub-cohorts, respectively). }
\label{fig:sensors}
\end{figure}

The analysis of sensor-based mobility (Figure \ref{fig:sensors}), available for subsets of those 70 subjects who met the analysis criteria, shows a differential pattern across the groups. During the Baseline and COVID periods, QoL-Resilient patients (n=15) had higher “running” step counts  per day (likely reflective of more intense activity instead of actual running) than the other two groups (Figure \ref{fig:sensors}a), while Pain-Susceptible (n=8) patients had the lowest (QoL vs PainS Mann Whitney, $p<0.003$ and $p<0.03$); in contrast, during COVID the three groups have almost identical counts, consistent with the mobility restrictions imposed by the pandemic. When using the accelerometer-based features, although the overall effective mobility was increased for all groups in the COVID period, the result is not statistically significant (Figure \ref{fig:sensors}b, Wilcoxon $p=0.37$). However, we observed that there is a significant difference in the Baseline period between QoL-Resilient (n=16) and ADL Susceptible (n=17) sub-cohorts (Mann Whitney, $p<0.05$, higher for QoL-Resilient patients). On the other hand, a subset of sleep sensor measurements, averaged across each period, in ADL-Susceptible (n=18) patients was affected the most between periods. We found that sleep onset latency (Figure \ref{fig:sensors}c) significantly increased during COVID(Mann Whitney $p<0.05$). In contrast, and the number of toss and turns (Figure \ref{fig:sensors}d) was consistently higher in the QoL group (n=10) with respect to the PainS group (n=8) across periods (Mann Whitney $p<0.05$ in both).

\subsection{Subcohorts do not differ in their responses to in-clinic questionnaires prior to the pandemic}
To identify possible a priori clinical markers of response, we examined scores on questionnaires collected in clinic during or before the Baseline period (Pre-Baseline) and responses to the COVID-Questions. There were no statistically significant differences across the 3 sub-cohorts in depression scores (BDI-II), pain catastrophizing tendencies (PCS), fear-avoidance behavior (FABQ), or overall disability (ODI) collected during or before the Baseline period (see supplemental material for statistics), suggesting the differences seen in EMA characteristics were not due to underlying personality differences or undiagnosed co-morbid mood conditions. Likewise, there were no differences seen across the 3 groups’ acute mental health scores reported during COVID-period or in the number of significant problems reported during COVID-period (in supplement), implying that all clusters were equally (albeit differently) stressed and concerned by the pandemic. 

\subsection{Subcohorts differ in their emotional reactions during the COVID-period and in their language qualities over time}
\begin{figure}[htbp]
\centerline{\includegraphics[scale=0.7]{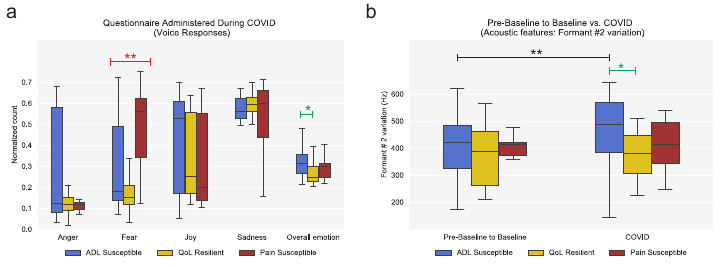}}
\caption{B{\bf oxplots of features from voice analysis.} Horizontal bars indicate significant differences using Kruskal Wallis test (red), Mann-Whitney (green) and Wilcoxon paired test (black). * and ** at the top of the bars indicates p$<$0.05 and p$<$0.01, respectively (a). NLP features captured from voice responses to a prompt administered during a one-week time frame in the COVID-Period show differences between the 3 sub-cohorts in emotional responses to the pandemic (N = 23, 15, and 10 patients answered this voice question for ADL Susceptible, QoL Resilient and Pain Susceptible sub-cohorts). (b) Variation of formant 2 (acoustic feature) were captured from voice responses, were averaged for both the Baseline period and Pandemic period and were compared between sub-cohorts (N = 21, 9, and 7) for ADL Susceptible, QoL Resilient and Pain Susceptible sub-cohorts).}
\label{fig:voice}
\end{figure}

While we did not find any evidence of underlying differences in emotional characteristics in our clusters based on in-clinic scores or mental health self-reports, we did find differences in psycholinguistic and acoustic properties of patients’ voice recordings during the study. Voice recordings were completed by 48/70 subjects during the COVID-Questions week, Baseline period, and Pandemic period. We show distinct differences between the sub-cohorts in the psycholinguistic content and acoustic speech properties of their speech.  Regarding acute responses to COVID Questions, QoL-Resilient patients showed significantly less use of fear-related words than the other cohorts, a finding that aligns with their EMA attributions (Figure \ref{fig:voice}a, Kolmogorov-Smirnov test, $p<0.01$). When looking at changes in speech across time, one of the most prominent acoustic features that differentiated both periods was variation in the frequency response of the acoustic resonance of the human vocal tract (i.e., formants, Figure \ref{fig:voice}b). We found that the interquartile range (IQR) of formant 2 (F2) values increased in the COVID-period for all patients (Wilcoxon, $p<0.02$), an effect which was most pronounced in the ADL-Susceptible sub-cohort (Wilcoxon, $p<0.005$). Differences in content-based features between periods were also informative. For example, the normalized count of sleep-related words was higher in the COVID-period for the ADL-Susceptible sub-cohort (Wilcoxon, $p<0.004$)). On the other hand, pain-related words were used less frequently in COVID-period for the QoL-Resilient sub-cohort (Wilcoxon, $p<0.004$).

Further details on the analysis can be found in the
\href{https://www.dropbox.com/s/i0du0cqhyee09h9/supplement_covid_ICDH.pdf?dl=0} {\tt Supplement}.

\section{Conclusions}
Following our initial hypotheses, we demonstrate multiple aspects of chronic pain patients’ everyday experiences were significantly altered by the COVID-19 pandemic, in ways that were not observed in the overall population, not related to pain intensity alone and not immediately clinically obvious. Instead, we show the emergence of 3 unique sub-cohorts of patients, some of whom were vulnerable to increased pain intensity during COVID (Pain-Susceptible), some to worsening activities of daily living (ADL-Susceptible), and others whose experiences largely \textit{improved} during the pandemic, including better mood, better sleep, and less medication utilization (QoL-Resilient). We did not know a priori which cluster a patient would be assigned to during the COVID-period; however, because we saw differences and changes pre-COVID as a function of sub-cohort, it may be possible for future work to predict how patients may respond to other stressors that they routinely experience. This work also highlights the utility of longitudinal multi-modal cross-environment digital analytics - in the future, these approaches could help ensure clinicians are well-informed when making clinical decisions across a broad range of domains and patient symptoms.

The patient groups were characterized by subtle but important differences in the interactions between their EMAs, how active they were prior to COVID, their sleep during COVID, and how they talked about their experiences during and prior to COVID, all of which suggest underlying differences in emotional and behavioral responses to stress (e.g., different coping mechanisms \cite{b18,b19} or different feelings of control versus helplessness \cite{b20,b21}). Importantly, the reported levels of pain intensity, while essential, by themselves would not have revealed the rich effect of the pandemic stress on the patients’ lives. The inclusion of multiple aspects of the pain experience – mood, sleep, medication use, and various activities – ultimately allowed for the dissociation of three unique patient responses, which together may have implications for differential, patient-centric remote therapy solutions or study engagement plans. That is not to say that all the features collected were necessary, nor that the more aspects of the chronic pain experience measured, the better the outcome (more data does not always mean better results). Instead, our results imply artificial intelligence techniques with multiple data collection mechanisms can provide deeper and important insights into patient experiences.  We show here that the relevant representational dimensionality of the chronic pain experience is not arbitrarily complex or large: as best exemplified by the MDS map, a limited number of properly defined factors may suffice to account for what could be considered a multi-modal chronic pain "index" representing the current state of the patient. This integrative representation would provide clinicians with a radically new approach to understanding chronic pain patients and perhaps enable them to proactively intervene in ways they may not have previously. Moreover, the introduction of digital technology allowed for the inclusion and assessment of novel, more naturalistic, and richer data streams not typically available via traditional research studies or patient care delivery. For example, mobility is a largely passive measure of utmost clinical importance which, outside of functional fitness assessments done at a visit, has otherwise only been estimated retrospectively and inaccurately \cite{b22}. Likewise, the limitation of yes/no or scale-based self-reported assessments can be enriched by allowing for more naturalistic language expression, bringing them closer to how in-clinic assessments are conducted (e.g., conversationally).

In particular, the sub-cohort we identified as Pain-Susceptible can additionally be characterized by having significantly less mobility during the Baseline period, and significantly higher preoccupation with fear-related topics during the COVID-period, both of which may offer additional opportunities for prevention (e.g., to incentivize or enable patients to improve mobility from the start) and intervention (e.g., to recommend concomitant psychological treatments to be used in conjunction with SCS to help reduce or reframe co-morbid anxious thoughts or limit fear-based actions/inactions).  We found a reduced amount of sleep hours and worse sleep quality for the ADL Susceptible sub-cohort in comparison with the other two groups, in line with the hypothesis that this sub-cohort required more time to fall asleep during COVID as measured by the under-the-mattress sleep sensor. We similarly observed alignment between self-reported activity and objective measures of activity – for example, our measure of effective mobility was always higher for the QoL Resilient sub-cohort than for ADL Susceptible group who struggled with various activities during COVID.  These findings might have clinical implications for prescribing sleep-related medication or recommending other sleep or mobility-based interventions, personalized based on patients' subcohort membership. 

In the case of acoustic features, an increase in variability (IQR) of F2 during the COVID period with respect to the Baseline period may indicate that patients expressed more negative emotions or had a more negative tone when answering the prompts \cite{b23,b24}. This pattern was most noticeable for the ADL Susceptible sub-cohort when comparing the Baseline period versus Pandemic period. Likewise, our extracted content features were also congruent with our sub-cohorts,  with QoL Resilient patients showing the lowest amount of sadness-related words during COVID compared to the other two groups. Our findings also suggest that there is a positive correlation between how frequently certain topics are mentioned and the effects of those topics on a patient’s life. For example, sleep-related words were used more frequently during the COVID period for the ADL Susceptible group, suggesting that sleep was affected for these patients (a finding that self-reported EMAs confirmed). Similarly, the use of pain-related words was reduced during COVID for the QoL resilient sub-cohort, which may indicate that pain intensity was not the main concern for these patients, also in line with their EMA ratings. This may be related to normal human negativity bias, the idea that negative events or problems loom larger in our memories and attention than positive moments \cite{b25,b26,b27,b28} and/or to the availability heuristic, where things that happen more frequently are more easily remembered or provoked \cite{b29} - as such, patients are more likely to talk about the things that bother them or that they are experiencing more often. 

Modern medicine is already being transformed by early applications of artificial intelligence and other analytic techniques (e.g., machine learning, deep learning).  The fields of radiology, cardiology, pathology, oncology, ophthalmology, and endocrinology are among those currently utilizing algorithms and systems to assist clinicians in detecting, diagnosing, and treating disease \cite{b30,b31,b32,b33}.  Interventional pain management clinicians are also uniquely poised to take advantage of these approaches for patients with chronic pain due not only to the chronic and subjective nature of pain but also the diverse types and large amounts of data needed to understand a patient more holistically. We have demonstrated that AI can detect relationships and changes in chronic pain patients’ experiences that were not clinically obvious prior to analysis and which may have been difficult for patients to notice or explain themselves.  AI allows for simplification of complex data and holds the promise to enable proactive intervention and aid physician decision making in real-time to improve outcomes beyond just pain intensity for patients with chronic pain. And while AI is expected to bring further innovation to medical care as its applications are developed and refined, it seems much more likely to help physicians than replace them \cite{b34,b35,b36}.

Our study has a number of limitations worth noting. One of them is the non-specificity of the questions administered. As the clinical study we investigated was launched well before the pandemic began, the data collected – including self-reported ratings, questions asked during COVID, and voice prompts - were not designed to specifically measure unexpected stress, barriers to on-going treatment, or attitudes about the pandemic.  As such, more work could be done to design and administer questions purposefully targeted to capture patient COVID-19 responses. We were also limited by our sample size, which again was largely driven by the fact that the pandemic was unexpected and many patients did not have sufficient data prior to the COVID period due to when they enrolled in the ENVISION study, which meant we could not reliably compare their pandemic data to an unaffected “control” time point; future analyses with a larger number of patients would be warranted and useful to test the robustness of the cluster assignments. Similarly, while we attempted to maximize the generalizability to all chronic pain patients by controlling for SCS status in the analysis, further study is needed given the very small amount of subjects without SCS in the study, and understanding the effects of starting, stopping, or maintaining SCS treatment including the choice of dates over which to analyze the data. The periods under consideration included months in the fall and winter seasons, and therefore potential seasonal effects may limit a strong causal interpretation between the observed differential responses and the pandemic stress (e.g., with respect to mobility and mood/emotional measures).  

While continuous digital health tracking and quantification is obviously applicable for “normal”, everyday occurrences, COVID-19 highlights its importance even more so, particularly in the realm of chronic pain monitoring and SCS treatment.  In fact, some research suggests that we could see an increase in chronic pain "after" the pandemic resulting from viral-associated corporeal damage, worsening pre-existing pain due to lack of treatment or aggravated physical and mental issues, and/or newly triggered pain due to exacerbated risk factors (e.g., economic stress, depression, inactivity, poor sleep, etc) \cite{b37}.  Moreover, it is also likely that co-morbid physical and mental health concerns \cite{b38}, as well as social concerns \cite{b39,b40}, will increase acutely or chronically as a consequence of the pandemic in a variety of patient cohorts (including chronic pain SCS patients).  
 
Understanding chronic pain patients more holistically  beyond traditional in-clinic pain score assessments alone is critical. Our findings have shown we can provide physicians with deep insight into their patients that may not have been previously possible by using advanced analytic techniques (e.g., machine learning) on dense, multi-dimensional data collected using minimally intrusive digital technologies. This research, and other work like it, has the potential to arm clinicians with meaningful insights outside of COVID-19 that can improve healthcare efficiency, therapy efficacy, and patient care delivery. Future direction of our on-going investigations will evaluate predicting comprehensive outcomes and stratification in patients with chronic pain for deep personalization of SCS and other therapies for clinical intervention \cite{b41}.

\section*{ \textsuperscript{**} Physician Author Group }
The NAVITAS and  ENVISION  Studies  Physician  Author  Group  includes  Richard  Rauck  (The  Center  for  Clinical  Research),  Eric  Loudermilk  (PCPMG  Clinical  Research  Unit),  Julio  Paez  (South  Lake  Pain  Institute),  Louis  Bojrab  (Forest  Health  Medical  Center),  John  Noles  (River  Cities  Interventional  Pain),  Todd  Turley  (Hope  Research  Institute),  Mohab  Ibrahim  (Banner  University  Medical  Center),  Amol  Patwardhan  (Banner  University  Medical  Center),  James  Scowcroft  (KC  Pain  Centers),  Rene  Przkora  (University  of  Florida),  Nathan  Miller  (Coastal  Pain  and  Spinal  Diagnostics),  and  Gassan  Chaiban  (Ochsner  Clinic  Foundation).

\section*{Acknowledgments}
The work presented here results from a collaboration between Boston Scientific and IBM Research. We thank Pritish Parida from IBM Research for his work in developing, calculating, and sharing the effective mobility metric used here. We would also like to thank all the patients that participated in this research and the entire team from the IBM-Boston Scientific joint research project for their work and dedication to this project and the larger parent studies.

\vspace{12pt}
\color{red}

\end{document}